\documentclass[lettersize,journal]{IEEEtran}
\usepackage{amsmath,amsfonts}
\usepackage{algorithmic}
\usepackage{algorithm}
\usepackage{array}
\usepackage[caption=false,font=normalsize,labelfont=sf,textfont=sf]{subfig}
\usepackage{textcomp}
\usepackage{stfloats}
\usepackage{url}
\usepackage{verbatim}
\usepackage{graphicx}
\usepackage{cite}
\usepackage{bm}
\usepackage{amssymb}
\usepackage{color}

\begin{document}

\title{Symbol-Level Precoding for Integrated Sensing and Communications: A Faster-Than-Nyquist Approach}

\author{Zihan Liao, Fan Liu,~\IEEEmembership{Member,~IEEE}
}



\maketitle

\begin{abstract}
In this paper, we propose a novel symbol-level precoding (SLP) method for a multi-user multi-input multi-output (MU-MIMO) downlink Integrated Sensing and Communications (ISAC) system based on faster-than-Nyquist (FTN) signaling. Our method minimizes the minimum mean squared error (MMSE) for target parameter estimation while guaranteeing per-user quality-of-service by exploiting constructive interference (CI) techniques. We tackle the non-convex problem using an efficient successive convex approximation (SCA) method. Numerical results demonstrate that our FTN-ISAC-SLP design significantly outperforms conventional benchmarks in both communication and sensing performance.
\end{abstract}


\section{Introduction}

\IEEEPARstart{I}{ntegrated} sensing and communications (ISAC) has been recognized as a key enabling technology for next-generation wireless networks (such as 5G-Advanced (5G-A) and 6G). It pursues a deep integration between wireless sensing and communication (S\&C) such that the two functionalities can be co-designed to improve the hardware-, spectral-, and energy-efficiency, as well as to acquire mutual performance gains \cite{pin2021integrated}. 


Numerous ISAC waveform design schemes can be broadly classified into two main approaches: non-overlapped resource allocation and fully unified waveform design. The former allocates orthogonal resources to S\&C to avoid interference, but suffers from resource inefficiency. The latter approach, which advocates shared wireless resources between S\&C, encompasses three schemes: Sensing-centric design (SCD), communication-centric design (CCD), and joint design (JD) \cite{liu2022integrated}. While SCD and CCD prioritize sensing or communication capabilities, JD schemes strive for a scalable trade-off between S\&C without relying on existing waveforms \cite{feng2020china}.

Although a tradeoff exists between S\&C performance, our goal is to achieve a substantial communication data rate without significantly compromising sensing performance. To this end, we introduce faster-than-Nyquist (FTN) signaling, a technique that improves symbol rate by reducing transmitted pulses temporally \cite{anderson2013faster}. However, FTN signaling violates the Nyquist criterion, resulting in inter-symbol interference (ISI). Alongside multi-user (MU) interference in MU-MIMO systems, interference occurs in both spatial and temporal domains. Traditional precoding designs mitigate interference through channel equalization techniques, such as zero forcing, but neglect the potential benefits of leveraging known interference at the ISAC base station (BS) \cite{spano2018faster}.

The inherent flexibility of the JD ISAC waveform design enables SLP implementation to exploit constructive interference (CI). In contrast to block-level precoding (BLP), which relies solely on channel state information (CSI), SLP designs transmit signals based on both the CSI and data information to manage interference constructively, thereby improving communication signal-to-interference-plus-noise ratio (SINR) \cite{li2020tutorial}. Consequently, FTN signaling and SLP become a perfect couple particularly suitable for ISAC applications, since the temporal interference can be utilized to enhance communication performance without impairing sensing performance, as evidenced in later sections of the paper.

In this paper, we propose a novel ISAC precoding technique referred to as FTN-ISAC-SLP. It merges the strategies discussed above, thus realizing performance augmentation for S\&C from both temporal and spatial dimensions. We first introduce the system model and performance metrics for the considered MU-MIMO ISAC system employing FTN signaling, and then formulate the ISAC precoding design into an symbol-level optimization problem. While the problem is non-convex in general, we propose a tailor-made successive convex approximation (SCA) method, which finds a near-optimal solution in polynomial time. Numerical results show that the proposed FTN-ISAC-SLP method achieves significant performance gain in terms of both S\&C compared to conventional Nyquist signaling and BLP approaches.

\section{System Model}
We consider a narrowband MIMO ISAC BS equipped with $N_t$ transmit antennas and $N_r$ receive antennas, which is serving $K$ downlink single-antenna users while detecting targets as a monostatic radar. Without loss of generality, we assume $K < N_t$. Before formulating the FTN-ISAC-SLP problem, we first elaborate on the system model and performance metrics of both radar sensing and communications.
\begin{figure}[h]
    \centering
    \includegraphics[scale=0.45]{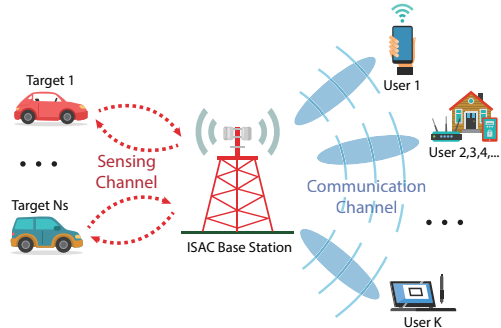}
    \vspace{-10pt}
    \caption{ISAC Downlink System.}
    \label{fig:system}
    \vspace{-10pt}
\end{figure}
\subsection{Signal Model}
Let $\mathbf{S}=[\mathbf{s}_0, \mathbf{s}_1, \cdots, \mathbf{s}_{K-1}]^{\top} \in \mathbb{C}^{K \times L}$ denote the symbol matrix to be transmitted, with $\mathbf{s}_k$ being the data stream intended for the $k$-th user with a block length $L$, and each entry being drawn from a given constellation. Unless otherwise specified, in this paper we consider a PSK constellation, since the extension to QAM constellations is straightforward \cite{li2020tutorial}. Moreover, let $\mathbf{X} = [\mathbf{x}_0, \mathbf{x}_1, \cdots, \mathbf{x}_{N_t-1}]^{\top} \in \mathbb{C}^{N_t \times L}$ be the precoded signal matrix, with $\mathbf{x}_n$ representing the data stream to be transmitted at the $n$-th antenna. Suppose that the precoded symbols $\mathbf{x}_i$ are passed through a root-raised-cosine (RRC) shaping filter $\varphi(t)$ with a roll-off factor $\alpha$ and a duration $T_0$. The band-limited signal is transmitted with an FTN-specific symbol interval $T = \tau T_0$ where $\tau \in [0,1]$. Under such a setting, the transmit FTN signal $x_n(t)$ can be expressed as
\begin{equation}
    x_n(t) = \sum_{i=0}^{L-1}x_{n,i}\varphi(t-iT),
\end{equation}
where $x_{n,i}$ represents the $i$-th element of $\mathbf{x}_n$.
\subsubsection{Communication Model}
Consider  the quasi-static flat fading channel matrix $\mathbf{H}_C = [\mathbf{h}_0^C,\mathbf{h}_1^C,\cdots,\mathbf{h}_{K-1}^C]^{\top}\in\mathbb{C}^{K \times N_t}$ that models the MUI among different data streams, where $\mathbf{h}_k^C$ is the channel vector of $k$-th user. Then the received signal at the $k$-th user can be expressed as
\begin{equation}
    r_k^C(t) = \sum_{i=0}^{N_t-1}h_{k,i}^Cx_i(t) + n_k^C(t),
\end{equation}
where $h_{k,i}^C$ represents the $i$-th elements of $\mathbf{h}_{C,k}$ and $n_k^C(t)$ is the complex-valued additive white Gaussian noise (AWGN) at the $k$-th user with zero mean and variance $\sigma_C^2$.

The received FTN signal after passing through a matched filter $\varphi^{*}(-t)$ at the $k$-th user is given by
\begin{equation}
\begin{aligned}
    y_k^C(t) &= r_k^C(t) * \varphi^{*}(-t) \\
    &= \sum_{i=0}^{N_t-1}h_{k,i}^Cx_i(t) * \varphi^{*}(-t) + n_k^C(t) * \varphi^{*}(-t) \\
    &= \sum_{i=0}^{N_t-1}\sum_{j=0}^{L-1}h_{k,i}^Cx_{i,j}\phi(t-jT) + \eta_k(t),
\end{aligned}
\end{equation}
where
\begin{equation}
\begin{aligned}
    \phi(t) &= \int_{-\infty}^{\infty}\varphi(\zeta)\varphi^{*}(\zeta-t)d\zeta, \\
    \eta_k(t) &= \int_{-\infty}^{\infty}n_k^C(\zeta)\varphi^{*}(\zeta-t)d\zeta.
\end{aligned}
\end{equation}
The $l$-th filtered sample at the $k$-th user $y_{k,l}^C = y_k^C(lT) \; (l = 0,1,\cdots,L-1)$ can be expressed as
\begin{equation}
\label{eq:filted_signal}
    y_{k,l}^C = \sum_{i=0}^{N_t-1}\sum_{j=0}^{L-1}h_{k,i}^Cx_{i,j}\phi((l-j)T) + \eta_k(lT),
\end{equation}
which can be written in a compact matrix form as
\begin{equation}
\label{eq:c_matrix_form_origin}
    \widetilde{\mathbf{Y}}_C = \mathbf{H}_C\mathbf{X}\mathbf{\Phi} + \widetilde{\mathbf{N}}_C,
\end{equation}
where $\widetilde{\mathbf{Y}}_C = (y_{k,l}) \in \mathbb{C}^{K \times L}$, and $\mathbf{\Phi} \in \mathbb{R}^{L \times L}$ is defined as
\begin{equation}
\label{eq:RGXH}
\begin{aligned}
    &\mathbf{\Phi} = \\&
    \begin{bmatrix}
        \phi(0) & \phi(-T) & \cdots & \phi(-(L-1)T) \\
        \phi(T) & \phi(0) & \cdots & \phi(-(L-2)T) \\
        \vdots & \vdots & \ddots & \vdots \\
        \phi((L-1)T) & \phi((L-2)T) & \cdots & \phi(0)
    \end{bmatrix},
\end{aligned}
\end{equation}
which is a positve semidefinite Toeplitz symmetric matrix. Moreover, $\widetilde{\mathbf{N}}_C = [\bm{\eta}_0, \bm{\eta}_1, \cdots, \bm{\eta}_{K-1}]^{\top}$, with $\bm{\eta}_k = [\eta_k(0), \eta_k(T), \cdots, \eta_k((L-1)T)]^\top$ being the corresponding noise vector at the $k$-th user. Notice that $\mathbb{E}[\bm{\eta}_k\bm{\eta}_k^H] = \sigma_C^2\mathbf{\Phi}$, which indicates that the noise received at each user is not independent. To decorrelate the noise, consider the eigenvalue decomposition of $\mathbf{\Phi}$ is $\mathbf{U}\mathbf{\Lambda}\mathbf{U}^{H}$ where $\mathbf{U}$ is a unitary matrix containing eigenvectors and $\mathbf{\Lambda}$ is a diagonal matrix composed by eigenvalues. Right-multiplying $\mathbf{U}$ at both sides of (\ref{eq:c_matrix_form_origin}) yields
\begin{equation}
\label{eq:c_matrix_form}
    \mathbf{Y}_C = \mathbf{H}_C\mathbf{X}\mathbf{U\Lambda} + \mathbf{N}_C,
\end{equation}
where $\mathbf{Y}_C=\widetilde{\mathbf{Y}}_C\mathbf{U}$ and $\mathbf{N}_C=\widetilde{\mathbf{N}}_C\mathbf{U}$. By doing so, the covariance matrix for row vectors of $\mathbf{N}_C$ becomes $\sigma_C^2\mathbf{\Lambda}$, i.e., a diagonal matrix.
\subsubsection{Radar Sensing Model}
Consider the target response matrix (TRM) $\mathbf{H}_R = [\mathbf{h}_0^R, \mathbf{h}_1^R, \cdots, \mathbf{h}_{N_r-1}^R]^{\top} \in \mathbb{C}^{N_r \times N_t}$ that models the sensing channel. Depending on the sensing scenarios, $\mathbf{H}_R$ can be of different forms. For example, in the case of an angular extended target model where all point-like scatterers are situated within the same range bin, we have
\begin{equation}
\mathbf{H}_R=\sum_{i=1}^{N_s}\alpha_i\mathbf{b}(\theta_i)\mathbf{a}^H(\theta_i),
\end{equation}
where $N_s$ is the number of scatterers, $\alpha_i$ and $\theta_i$ denote the reflection coefficient and the angle of the $i$-th scatterer, and $\mathbf{a}\left(\theta\right) \in \mathbb{C}^{N_t \times 1}$ and $\mathbf{b}\left(\theta\right) \in \mathbb{C}^{N_r \times 1}$ are transmit and receive steering vectors. 

However, the BS lacks prior knowledge of scatterer numbers and angles, and more advanced radar channel models may be applicable. As a result, there is typically no definitive structure for the sensing channel model. To ensure generality, following \cite{liu2022integrated}, we consider a generic TRM $\mathbf{H}_R$ instead of a specific model. Subsequently, we may extract the parameters of the sensing channel from the estimated TRM.

Similar to the communication model, $y_{k,l}^R=y_k^R(lT)(l=0,1,\cdots,L-1)$, the $l$-th sample of the received echo signal at the $k$-th receive antenna, can be written as
\begin{equation}
\begin{aligned}
    y_{k,l}^R &= y_k^R(lT) = \sum_{i=0}^{N_t-1}h_{k,i}^Rx_i(lT) + n_k^R(lT) \\
    &= \sum_{i=0}^{N_t-1}\sum_{j=0}^{L-1}h_{k,i}^Rx_{i,j}\varphi((l-j)T) + n_k^R(lT),
\end{aligned}
\end{equation}
where $h_{k,i}^R$ represents the $i$-th elements of $\mathbf{h}_{R,k}$ and $n_k^R(t)$ is the complex-valued AWGN at the $k$-th receive antenna with zero mean and variance $\sigma_R^2$. At the sensing receiver, we directly sample the received signal without passing it through the pulse-shaping filter, yielding the following radar received signal model
\begin{equation}
\label{eq:radar}
    \mathbf{Y}_R = \mathbf{H}_R\mathbf{X}\mathbf{C}^{\top} + \mathbf{N}_R,
\end{equation}
where $\mathbf{N}_R$ denotes an AWGN matrix, with zero mean and the variance of each entry being $\sigma_R^2$. Here we assume $\mathbf{h}=\mathrm{vec}(\mathbf{H}_R)\thicksim\mathcal{CN}(\bm{\mu}_\mathbf{h},\sigma_H^2\mathbf{I})$ and $\mathbf{C}$ is given by
\begin{equation}
\label{eq:conv_matrix}
    \mathbf{C} =
    \begin{bmatrix}
        c_0 & 0 & \cdots & 0 \\
        c_1 & c_0 & \ddots & \vdots \\
        \vdots & \vdots & \ddots & \vdots \\
        c_{P-1} & c_{P-2} & \ddots & 0 \\
        0 & c_{P-1} & \ddots & c_0 \\
        \vdots & \vdots & \vdots & \vdots \\
        0 & \cdots & \cdots & c_{P-1}
    \end{bmatrix},
\end{equation}
where $\mathbf{c}=[c_0,c_1,\cdots,c_{P-1}]^{\top}$ is the sample vector for RRC function $\varphi(t)$ with sampling number $P$.

\textit{Remark:} In the communication model we attempt to detect the signal $\mathbf{S}$ from $\mathbf{X}$ in the receiver side, thus we pass the received signal to RRC matched filter to maximize the received SINR for each precoded symbol. In the sensing model our aim is to recover the TRM $\mathbf{H}_R$ from the raw observation (\ref{eq:radar}), rather than to recover $\mathbf{X}$. Therefore, we treat $\mathbf{X}\mathbf{C}^{\top}$ as an equivalent transmitted waveform and regard (\ref{eq:radar}) as the sufficient statistics for estimating $\mathbf{H}_R$, which needs not to be match-filtered by the RRC pulse.
\vspace{-10pt}

\subsection{CI Constraint for Communication}
CI constraint represents the constraint that pushes the received symbols away from all of their corresponding detection thresholds within the modulated-symbol constellation, thereby contributing positively to the overall useful signal power.
Following the CI expression given in \cite{masouros2015exploiting}, for any transmitted symbol $s$ and its corresponding received symbol $y$ at the receiver side, they must satisfy the following inequality to exploit the CI effect
\begin{equation}
    \left| \Im\left\{s^{*}y\right\} \right| - \Re\left\{s^{*}y\right\}\tan\theta \leq -\sqrt{\Gamma\sigma^2}\tan\theta,
\end{equation}
where $\sigma^2$ is the variance of the noise imposes on this symbol, $\Gamma$ is the required SNR and $\theta$ depends on the modulation type.
Let $\mathbf{Y}_C=[\mathbf{y}_0,\mathbf{y}_1,\cdots,\mathbf{y}_{K-1}]^{\top}$ and $\bm{\sigma}=\sqrt{\mathrm{diag}(\sigma_C^2\mathbf{\Lambda})}=[\sigma_C\Lambda_{0,0},\sigma_C\Lambda_{1,1},\cdots,\sigma_C\Lambda_{L-1,L-1}]^{\top}$ where $\mathrm{diag}$ refers to the operation taking the entries in the diagonal and stacking them as a vector. Then the CI constraint for the $k$-th user can be written as 
\begin{equation}
\label{eq:comlex_ci_origin}
    \left| \Im\left\{\mathbf{s_k^{*}} \circ \mathbf{y}_k\right\} \right| - \Re\left\{\mathbf{s_k^{*}} \circ \mathbf{y}_k\right\}\tan\theta \leq (-\sqrt{\Gamma_k}\tan\theta)\bm{\sigma}, \; \forall{k},
\end{equation}
where $\Gamma_k$ is the required SINR of the $k$-th user, and $\circ$ refers to the Hadamard product.
For the received symbol at the $k$-th user we have
\begin{equation}
    \mathbf{y}_k^{\top} = {\mathbf{h}_k^C}^{\top}\mathbf{XU\Lambda}.
\end{equation}
By noting the fact that $\mathbf{s_k^{*}} \circ \mathbf{y}_k$ can be equivalently expressed as $\mathbf{s_k^{*}} \circ \mathbf{y}_k = \mathbf{S}_k^{*}\mathbf{y}_k$, where
\begin{equation}
    \mathbf{S}_k = \mathrm{Diag}(\mathbf{s}_k) =
    \begin{bmatrix}
        e^{j\phi_{k,0}} & 0 & \cdots & 0 \\
        0 & e^{j\phi_{k,1}} & \cdots & 0 \\
        \vdots & \vdots & \ddots & \vdots \\
         0 & 0 & \cdots & e^{j\phi_{k,L-1}}
    \end{bmatrix},
\end{equation}
the inequality (\ref{eq:comlex_ci_origin}) can be recast to
\begin{equation}
\label{eq:comlex_ci}
\begin{aligned}
    \left| \Im\left\{{\mathbf{h}_k^C}^{\top}\mathbf{XU\Lambda}\mathbf{S}^{*}_k\right\} \right| - \Re\left\{{\mathbf{h}_k^C}^{\top}\mathbf{XU\Lambda}\mathbf{S}^{*}_k\right\}\tan\theta \\
    \leq (-\sqrt{\Gamma_k}\tan\theta)\bm{\sigma}^{\top}, \; \forall k,
\end{aligned}
\end{equation}
which is a linear constraint in $\mathbf{X}$ and is thus convex.
\vspace{-10pt}

\subsection{MMSE for Radar Sensing}
Let $\mathbf{y}_R=\mathrm{vec}(\mathbf{Y}_R)$, $\mathbf{h}_R=\mathrm{vec}(\mathbf{H}_R)$ and $\mathbf{n}_R=\mathrm{vec}(\mathbf{N}_R)$. Equation (\ref{eq:radar}) can be expanded as
\begin{equation}
    \mathbf{y}_R=(\mathbf{C}\mathbf{X}^{\top}\otimes\mathbf{I}_{N_r})\mathbf{h}_R + \mathbf{n}_R.
\end{equation}
According to \cite{kay1993fundamentals}, the corresponding MMSE for estimating $\mathbf{h}_R$ from the noisy observation $\mathbf{y}_R$ is
\begin{equation}
\begin{aligned}
    \mathrm{MMSE} &= \mathbb{E}(\Vert \mathbf{h}_R - \mathbf{h}_{R}^\mathrm{MMSE} \Vert_2^2) \\
    &= \mathrm{tr}\left(\left(\sigma_H^{-2}\mathbf{I}+\sigma_R^{-2}(\mathbf{C}\mathbf{X}^{\top}\otimes\mathbf{I}_{N_r})^{H}(\mathbf{C}\mathbf{X}^{\top}\otimes\mathbf{I}_{N_r})\right)^{-1}\right) \\
    &=
    \mathrm{tr}\left(\left(\sigma_H^{-2}\mathbf{I}+\sigma_R^{-2}(\mathbf{X}^{*}\mathbf{C}^{\top}\mathbf{C}\mathbf{X}^{\top}\otimes\mathbf{I}_{N_r})\right)^{-1}\right) \\
    &=
    \sigma_R^{2}N_r\mathrm{tr}\left(\left(\frac{\sigma_R^2}{\sigma_H^2}\mathbf{I}+\mathbf{X}\mathbf{\Psi}\mathbf{X}^{H}\right)^{-1}\right),
\end{aligned}
\end{equation}
where $\mathbf{\Psi}=\mathbf{C}^{\top}\mathbf{C}$. Notice that this expression is non-convex in $\mathbf{X}$, which will be tackled in the next section.

\section{FTN-ISAC Symbol-Level Precoding}
\subsection{Problem Fomulation}
Based on the discussion above, the precoding optimization
problem can be expressed as
\begin{equation}
\label{opt:origin_mimo}
\begin{aligned}
    \underset{\mathbf{X}}{\mathrm{min}} &\;
    f(\mathbf{X})=\mathrm{tr}\left(\left(\frac{\sigma_R^2}{\sigma_H^2}\mathbf{I}+\mathbf{X}\mathbf{\Psi}\mathbf{X}^{H}\right)^{-1}\right) \\
    s.t. &\;
    \left| \Im\left\{{\mathbf{h}_k^C}^{\top}\mathbf{XU\Lambda}\mathbf{S}^{*}_k\right\} \right| - \Re\left\{{\mathbf{h}_k^C}^{\top}\mathbf{XU\Lambda}\mathbf{S}^{*}_k\right\}\tan\theta \\ &\qquad\qquad\qquad\qquad\qquad
    \leq (-\sqrt{\Gamma_k}\tan\theta)\bm{\sigma}^{\top}, \; \forall k, \\
    &\; \Vert\mathbf{X}\mathbf{C}^{\top}\Vert_F^2 \leq E.
\end{aligned}
\end{equation}
That is, we design the precoded symbol matrix $\mathbf{X}$ for the to-be-transmitted symbol matrix $\mathbf{S}$, such that the MMSE for radar sensing is minimized while guaranteeing the CI conditions for communication under a given energy budget $E$.
\vspace{-5pt}

\subsection{Lower-Bound for the MMSE}
We first derive the lower bound of problem (\ref{opt:origin_mimo}) by considering the following optimization problem that solely minimizes the MMSE without imposing CI constraints
\begin{equation}
\label{opt:lower_bound}
\begin{aligned}
    \underset{\mathbf{X}}{\mathrm{min}} &\;
    f(\mathbf{X})=\mathrm{tr}\left(\left(\frac{\sigma_R^2}{\sigma_H^2}\mathbf{I}+\mathbf{X}\mathbf{\Psi}\mathbf{X}^{H}\right)^{-1}\right) \;
    s.t. \Vert\mathbf{X}\mathbf{C}^{\top}\Vert_F^2 \leq E.
\end{aligned}
\end{equation}
By letting $\widetilde{\mathbf{X}}=\mathbf{X}\mathbf{C}^{\top}$, problem (\ref{opt:lower_bound}) can be recast to 
\begin{equation}
\label{opt:equiv_lower_bound}
\begin{aligned}
    \underset{\widetilde{\mathbf{X}}}{\mathrm{min}} &\;
    f(\widetilde{\mathbf{X}})=\mathrm{tr}\left(\left(\frac{\sigma_R^2}{\sigma_H^2}\mathbf{I}+\widetilde{\mathbf{X}}\widetilde{\mathbf{X}}^{H}\right)^{-1}\right) \;
    s.t. \Vert\widetilde{\mathbf{X}}\Vert_F^2 \leq E
\end{aligned}
\end{equation}
which is independent of $\tau$. According to \cite{yang2007mimo}, the optimal value of problem (\ref{opt:equiv_lower_bound}) is
\begin{equation}
    f_{\min}=\sum_{i=1}^{N_t}\frac{1}{\lambda_i^2+\frac{\sigma_R^2}{\sigma_H^2}},
\end{equation}
where $\lambda_i^2=\left(\kappa-\frac{\sigma_R^2}{\sigma_H^2}\right)^+$ and $a^+=\max(a,0)$. The constant $\kappa$ is chosen to satisfy $\sum_{i=1}^{N_t}\lambda_i^2=E$.
\vspace{-5pt}

\subsection{SCA Algorithm for Solving Problem (\ref{opt:origin_mimo})}
It can be readily observed that the feasible region $\mathcal{Q}$ for (\ref{opt:origin_mimo}) is convex, whereas the objective function is not. To address this issue, we propose an SCA algorithm to solve problem (\ref{opt:origin_mimo}) in an iterative manner.

To proceed with the SCA algorithm, we approximate $f(\mathbf{X})$ using its first-order Taylor expansion near a
given point $\mathbf{X}_i\in\mathcal{Q}$ as
\vspace{-2pt}
\begin{equation}
    f(\mathbf{X}) \approx f(\mathbf{X}_i) + \Re\left\{\mathrm{tr}\left(\nabla f(\mathbf{X}_i)^{H}(\mathbf{X}-\mathbf{X}_i)\right)\right\}
\end{equation}
\vspace{-2pt}
where $\nabla f(\cdot)$ represents the gradient of $f(\cdot)$ and $\nabla f(\mathbf{X}_i)$ can be calculated as
\begin{equation}
\label{eq:gradient}
\begin{aligned}
    &\nabla f(\mathbf{X}_i) = \\&
    -2\left(\frac{\sigma_R^2}{\sigma_H^2}\mathbf{I}+\mathbf{X}_i\mathbf{\Psi}\mathbf{X}_i^{H}\right)^{-1}\left(\frac{\sigma_R^2}{\sigma_H^2}\mathbf{I}+\mathbf{X}_i\mathbf{\Psi}\mathbf{X}_i^{H}\right)^{-1}\mathbf{X}_i\mathbf{\Psi}
\end{aligned}
\end{equation}
At the $(i+1)$-th iteration of the SCA algorithm, we solve the following convex optimization problem
\begin{equation}
\label{opt:sca_mimo}
\begin{aligned}
    \underset{\mathbf{X}}{\mathrm{min}} &\;
    g(\mathbf{X})=\Re\left\{\mathrm{tr}\left(\nabla f(\mathbf{X}_i)^{H}(\mathbf{X}-\mathbf{X}_i)\right)\right\} \\
    s.t. &\;
    \left| \Im\left\{{\mathbf{h}_k^C}^{\top}\mathbf{XU\Lambda}\mathbf{S}^{*}_k\right\} \right| - \Re\left\{{\mathbf{h}_k^C}^{\top}\mathbf{XU\Lambda}\mathbf{S}^{*}_k\right\}\tan\theta \\ &\qquad\qquad\qquad\qquad\qquad
    \leq (-\sqrt{\Gamma_k}\tan\theta)\bm{\sigma}^{\top}, \; \forall k, \\
    &\; \Vert\mathbf{X}\mathbf{C}^{\top}\Vert_F^2 \leq E,
\end{aligned}
\end{equation}
where $\mathbf{X}_i\in\mathcal{Q}$ is the $i$-th iterative point. By solving the convex problem (\ref{opt:sca_mimo}), we obtain an optimal solution $\mathbf{X}^{\star}\in\mathcal{Q}$. Observing $g(\mathbf{X}^{\star})\leq 0$, it follows that $f(\mathbf{X}^{\star})\leq f(\mathbf{X}_i)$ when $\mathbf{X}^{\star}$ is close to $\mathbf{X}_i$ and linear approximation holds, indicating $\mathbf{X}^{\star}-\mathbf{X}_i$ is a descent direction. This idea is similar to gradient descent, as we step against the gradient direction, minimizing $g(\mathbf{X})$ which is the projection along the gradient, while keeping the next step within the feasible region by exploiting the convexity.

With a properly chosen step size $t\in[0,1]$, one may get the $(i+1)$-th iteration point as
\begin{equation}
    \mathbf{X}_{i+1}=\mathbf{X}_i+t(\mathbf{X}^{\star}-\mathbf{X}_i)=(1-t)\mathbf{X}_i+t\mathbf{X}^{\star}.
\end{equation}
Since $\mathbf{X}_i,\mathbf{X}^{\star} \in \mathcal{Q}$ by the definition of convexity, we have $\mathbf{X}_{i+1} \in \mathcal{Q}$, which is a feasible solution to problem (\ref{opt:origin_mimo}).

We are now ready to present Algorithm \ref{alg:sca} to solve problem (\ref{opt:sca_mimo}) based on the discussion above.
\begin{algorithm}
\caption{SCA Algorithm for Solving (\ref{opt:origin_mimo})}
\label{alg:sca}
\begin{algorithmic}[1]
    \REQUIRE $\mathbf{H}$, $\mathbf{\Psi}$, $E$, $\mathbf{S}$, $\bm{\sigma}$, $\Gamma_k,\forall k$, the execution threshold $\epsilon$ and the maximum iteration number $i_{\max}$.
    \ENSURE $\mathbf{X}^{\star}$
    \STATE {initialize $\mathbf{X}_0\in\mathcal{Q}$ by picking up $\mathbf{X}_{-1}$ randomly and solving problem (\ref{opt:sca_mimo}), $i=0$.}
    \REPEAT{
        \STATE Calculate the gradient $\nabla f(\mathbf{X}_i)$ by equation (\ref{eq:gradient}).
        \STATE Solve problem (\ref{opt:sca_mimo}) to obtain $\mathbf{X}^{\star}$.
        \STATE {
            Update the solution by
            $\mathbf{X}_{i+1} = \mathbf{X}_i + t\left(\mathbf{X}^{\star}-\mathbf{X}_i\right)$,
            where $t$ is determined by using the exact line search.
        }
        \STATE $i=i+1$.
    }
    \UNTIL {
        $\Vert \mathbf{X}_i-\mathbf{X}_{i-1}\Vert_F^2\leq\epsilon$ or $i=i_{\max}$.
    }
    \STATE $\mathbf{X}^{\star}=\mathbf{X}_i$
\end{algorithmic}
\end{algorithm}

\section{Numerical Results}

\begin{figure*}[t]
	\centering
	\begin{minipage}{0.48\linewidth}
		\centering
            \includegraphics[scale=0.485]{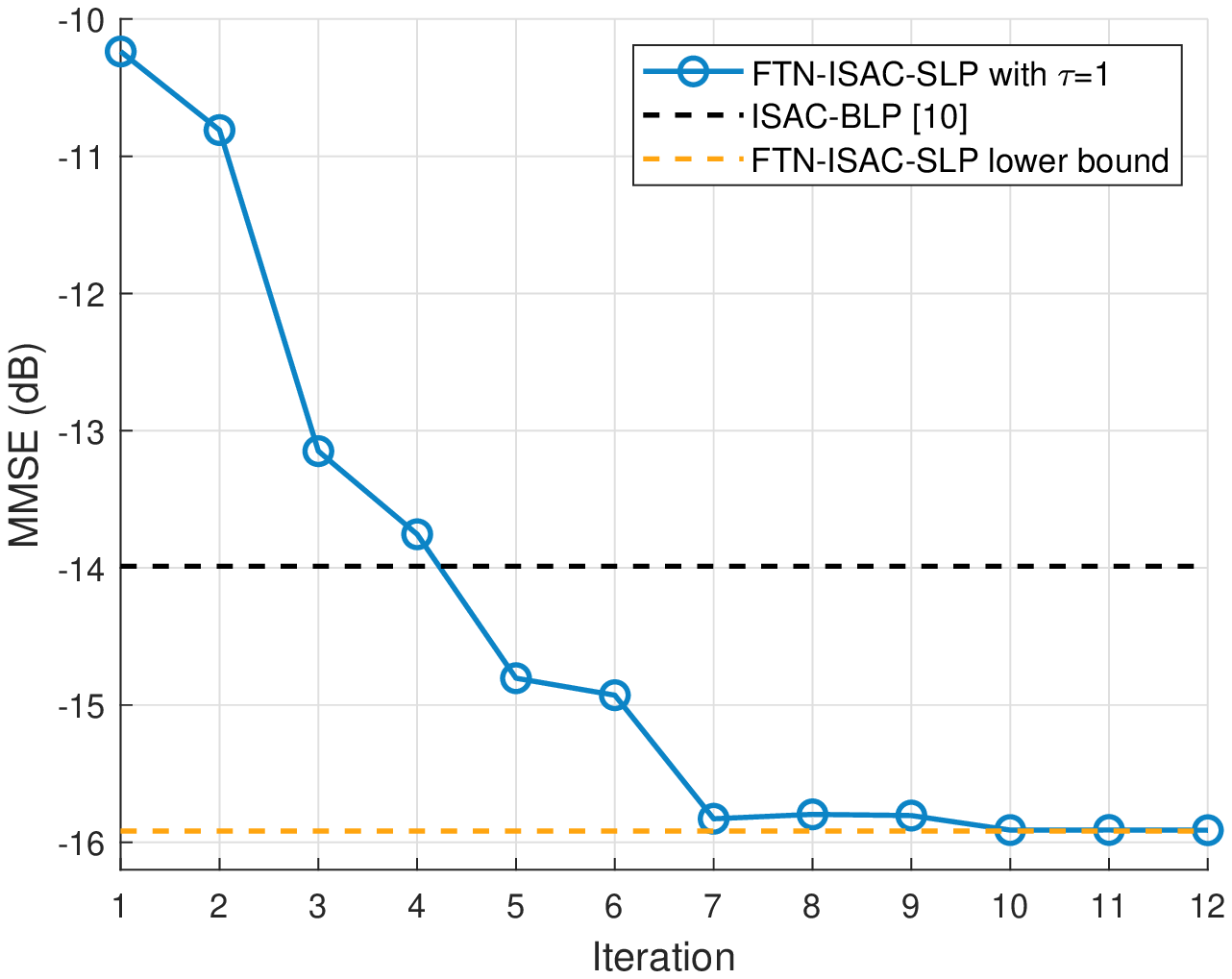}
            \vspace{-10pt}
            \caption{MMSE versus SCA iteration in case of $K=12$, $\Gamma=15\;\mathrm{dB}$, $E = 40\;\mathrm{dBm}$.}
            \label{fig:iter_mmse}
	\end{minipage}\qquad
	\begin{minipage}{0.48\linewidth}
		\centering
            \includegraphics[scale=0.485]{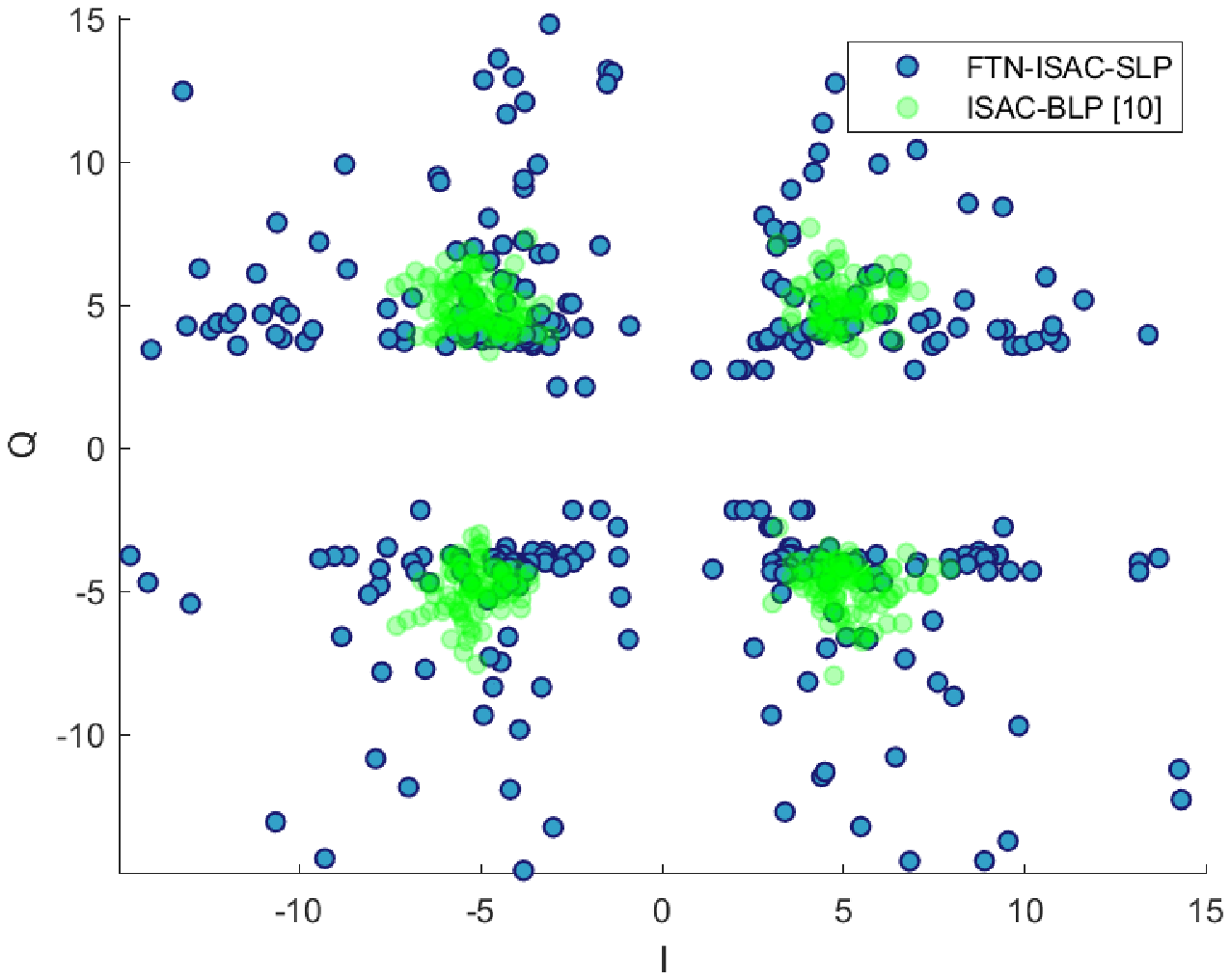}
            \vspace{-10pt}
            \caption{Constellation plot of the received symbols in case of $K=12$, $\Gamma=15\;\mathrm{dB}$, $E=35\;\mathrm{dBm}$.}
            \label{fig:constellation}
	\end{minipage}
	\begin{minipage}{0.48\linewidth}
            \centering
            \includegraphics[scale=0.485]{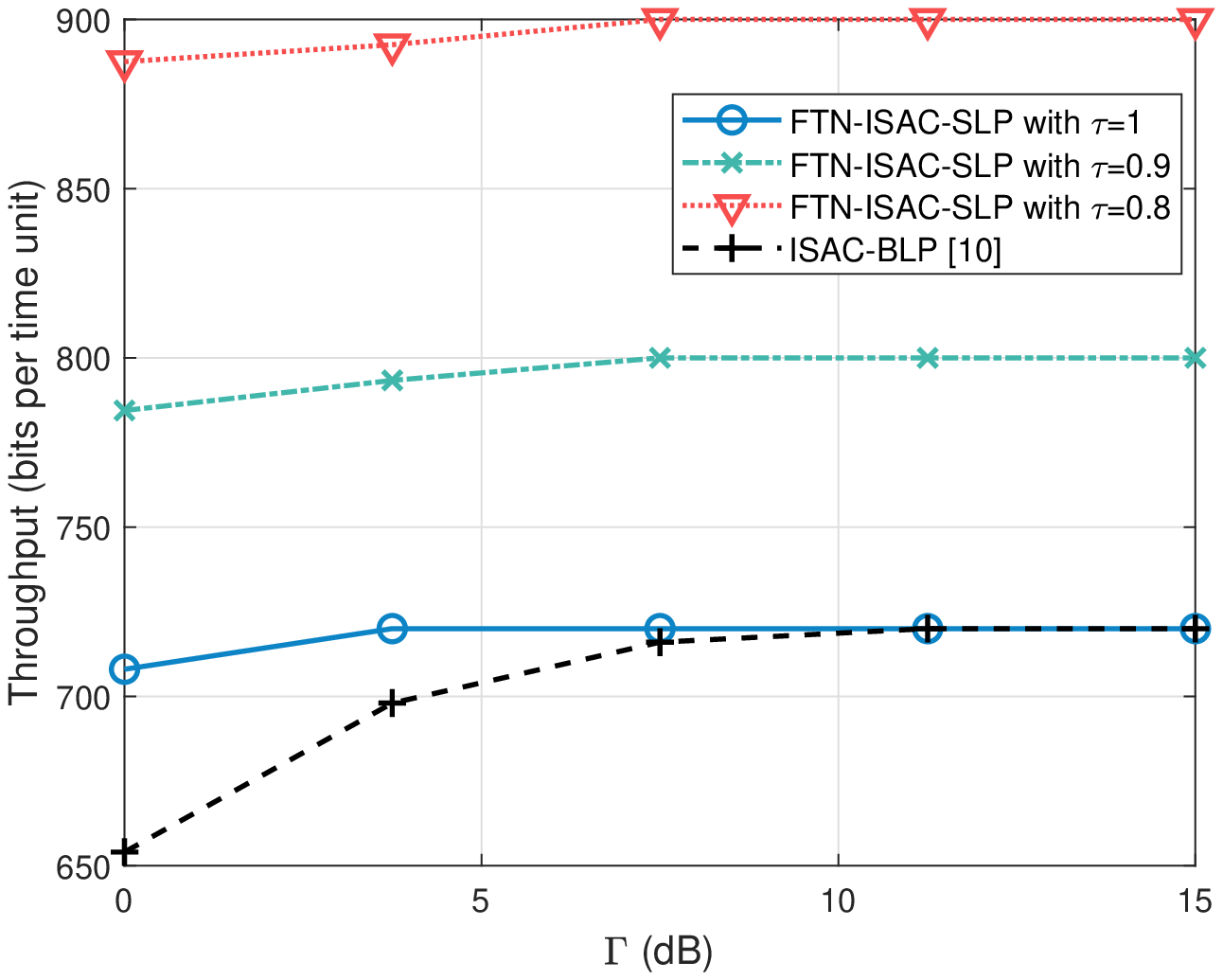}
            \vspace{-10pt}
            \caption{Throughput versus SNR in the case of $K=12$, $E=40\;\mathrm{dBm}$.}
            \label{fig:tp_sinr}
	\end{minipage}\qquad
	\begin{minipage}{0.48\linewidth}
            \centering
            \includegraphics[scale=0.49]{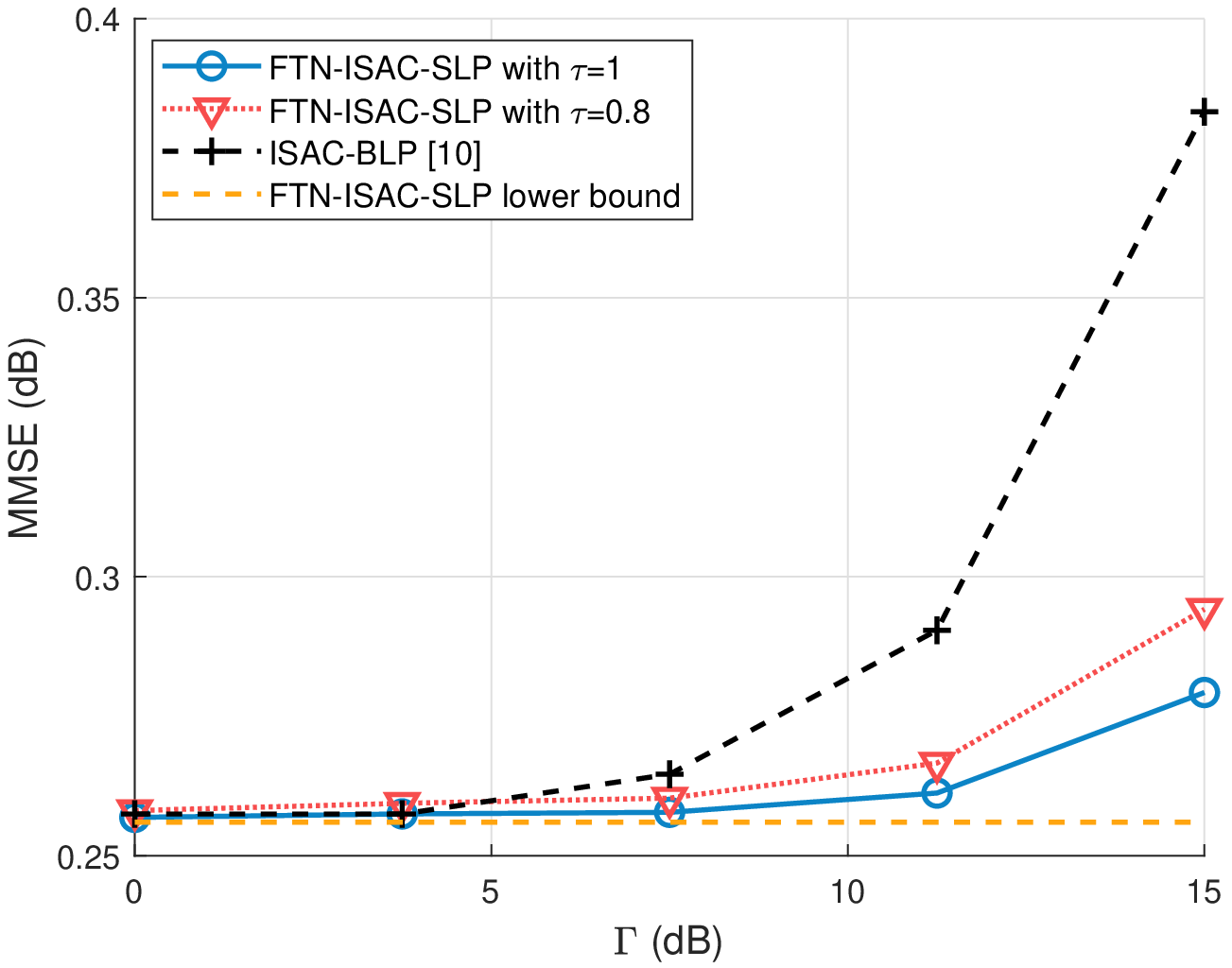}
            \vspace{-10pt}
            \caption{MMSE versus SNR, in the case of $K=8$, $E = 30\;\mathrm{dBm}$.} 
            \label{fig:mmse_sinr}
	\end{minipage}
\end{figure*}

In this section, we provide numerical results to verify the superiority of the proposed FTN-ISAC-SLP approaches. Without loss of generality, we consider an ISAC BS that is equipped with $N_t = 16$ and $N_r = 20$ antennas for its transmitter and receiver. The noise variances are set as $\sigma^2_C = \sigma^2_R = 0\;\mathrm{dBm}$, and the frames length is set as $L = 30$. The quantity of $\bm{\mu}_\mathbf{h}$ has minimal impact on the optimization discussed in this paper and is therefore set to $\mathbf{0}$. The variance of TRM fluctuations is set as $\sigma^2_H=20\;\mathrm{dBm}$, with each element of $\mathbf{H}_R$ drawn from $\mathcal{CN}(0,\sigma_H^2)$. Symbol duration $T_0$ is set to $1\;\mathrm{ms}$. We also assume a Rayleigh fading communication channel, with each element of $\mathbf{H}_C$ independently drawn from $\mathcal{CN}(0,\sigma_C^2)$. Without loss of generality, all the communication users are imposed with the same worst-case QoS, i.e., $\Gamma_k=\Gamma,\forall k$.

Our baseline ISAC-BLP is the ISAC beamforming method from \cite{liu2021cramer}, namely the block-level precoding method to find the optimal linear precoding matrix $\mathbf{W}_{DF}$ that minimizes the sensing CRB with guaranteed per-user SINR, through solving the problem below.
\begin{equation}
\label{opt:isac-blp}
\begin{aligned}
    \underset{\mathbf{W}_{DF}}{\mathrm{min}} &\; \mathrm{MMSE}(\mathbf{W}_{DF}) = 
    \frac{\sigma_R^2N_r}{L}\left(\left(\frac{\sigma_R^2}{\sigma_H^2}\mathbf{I}+\mathbf{W}_{DF}\mathbf{W}_{DF}^{H}\right)^{-1}\right) \\
    s.t. &\;
    \gamma_k \geq \Gamma_k, \; \forall k, \; L\Vert\mathbf{W}_{DF}\Vert_F^2 \leq E.
\end{aligned}
\end{equation}
where $\gamma_k$ is the SINR at $k$-th user. In order to ensure a fair comparison, we replace the original objective function CRB in \cite{liu2021cramer} with the MMSE.

Fig. \ref{fig:iter_mmse} shows the convergence performance of the proposed SCA algorithm. The tolerance threshold of the algorithm is set as $\epsilon=10^{-4}$. The algorithm converges and approaches to the lower bound we derived in above section. It can be observed that the proposed FTN-ISAC-SLP method outperforms the benchmark block-level design.

Fig. \ref{fig:constellation} shows the constellation plots for both the ISAC-BLP and FTN-ISAC-SLP approaches. The green points depict the region for SINR constraint while the blue points depict the region for CI constraint. It is clearly observed that the resulting CI constellation genearlly yields larger SNR compared to the block-level precoding.

In Fig. \ref{fig:tp_sinr}, we show the communication throughput performance with increased SNR threshold. Suppose the number of successfully recovered bits is $N_b$, the throughput is calculated by $N_b/\tau$ per time unit. We set FTN duration factor $\tau=0.8,0.9$ and $1$ respectively, and for ISAC-BLP $\tau=1$. As $\tau$ decreases, the throughput increases. When $\tau=1$, $T=T_0$, the FTN signaling reduces to Nyquist pulse shaping, whereas it still outperforms the ISAC-BLP method thanks to the exploitation of the CI effect.

Finally in Fig. \ref{fig:mmse_sinr}, we show the radar estimation MMSE with increased SNR threshold for communication users. It is observed that when the communications SNR is on the rise, the estimation performance becomes worse, which indicates that there is an inherent tradeoff between communication and sensing performance.  An increasing trend of MMSE when $\tau$ increases is observed, because the power constraint $\Vert\mathbf{X}\mathbf{C}^{\top}\Vert_F^2\leq E$ is tightened as $\tau$ increases. Again, our results show the superiority of the proposed FTN-ISAC-SLP method over that of the ISAC-BLP due to leveraging the CI constraint.
\vspace{-10pt}

\section{Conclusion}
This paper studied symbol-level precoding for faster-than-Nyquist signaling in ISAC, where a precoded symbol matrix is developed to carry out target sensing and information signaling simultaneously. In particular, we guarantee the per-user constructive interference constraint in the downlink while minimizing the MMSE for target estimation. Despite the non-convexity of the formulated precoding problem, we design an effective successive convex approximation method, which, at each iteration, resolves a second-order cone program subproblem. The superiority of the proposed FTN-ISAC-SLP method is demonstrated by numerical results, which show that our method is capable of greatly enhancing both communication and sensing performance compared to conventional block-level precoding based on Nyquist pulse shaping.

\bibliographystyle{IEEEtran}
\bibliography{reference}

\end{document}